\documentclass[aps,prL,twocolumn,superscriptaddress,showpacs,floatfix]{revtex4}
\usepackage{graphicx}
\usepackage{latexsym}
\usepackage{stmaryrd}
\usepackage{amsmath}
\usepackage{amssymb}
\usepackage{epstopdf}
\usepackage[colorlinks,linkcolor=red,anchorcolor=red,citecolor=blue]{hyperref}

\begin{document}
\makeatletter
\newcommand{\rmnum}[1]{\romannumeral #1}
\newcommand{\Rmnum}[1]{\expandafter\@slowromancap\romannumeral #1@}
\makeatother

\title{Measurement of the Chern Number for Non-Hermitian Chern Insulators}

\author{Hongfang Liu }
\affiliation{School of Physical Science and Technology, Soochow University, Suzhou, 215006, China}
\affiliation{Institute for Advanced Study, Soochow University, Suzhou 215006, China}
\affiliation{Jiangsu Key Laboratory of Thin Films, Soochow University, Suzhou 215006, China}
\author{Ming Lu }
\affiliation{Beijing Academy of Quantum Information Sciences, Beijing 100193, China}
\author{Shengdu Chai }
\affiliation{Department of Physics, Fudan University, Shanghai 200433, China}
\affiliation{Institute for Nanoelectronic Devices and Quantum Computing, Fudan University, Shanghai 200433, China}
\author{Zhi-Qiang Zhang }\email{zhangzhiqiangphy@163.com}
\affiliation{School of Physical Science and Technology, Soochow University, Suzhou, 215006, China}
\affiliation{Institute for Advanced Study, Soochow University, Suzhou 215006, China}
\author{Hua Jiang }\email{ jianghuaphy@fudan.edu.cn}
\affiliation{Institute for Advanced Study, Soochow University, Suzhou 215006, China}
\affiliation{Institute for Nanoelectronic Devices and Quantum Computing, Fudan University, Shanghai 200433, China}

\date{\today}

\begin{abstract}
The identification of the topological invariant of a topological system is crucial in experiments. However, due to the inherent non-Hermitian features, such determination is notably challenging in non-Hermitian systems. Here, we propose that the magnetic effect can be utilized to measure the Chern number of the non-Hermitian Chern insulator. We find that the splitting of non-Hermitian bands under the magnetic field is Chern number dependent. Consequently, one can easily identify the Chern number by analyzing these splitting sub-bands. From the experimental perspective, the measurement of non-Hermitian bands is demonstrated in LC electric circuits. Furthermore, we find that the non-Hermiticity can drive open (closed) orbits of sub-bands in the Hermitian limit closed (open), which can also be identified by our proposal. These phenomena highlight the distinctive capabilities of non-Hermitian systems. Our results facilitate the detection of Chern numbers for non-Hermitian systems and may motivate further studies of their topological properties.
\end{abstract}

\maketitle

\section{Introduction}\label{S1}
\begin{figure}
\includegraphics[width=0.48\textwidth]{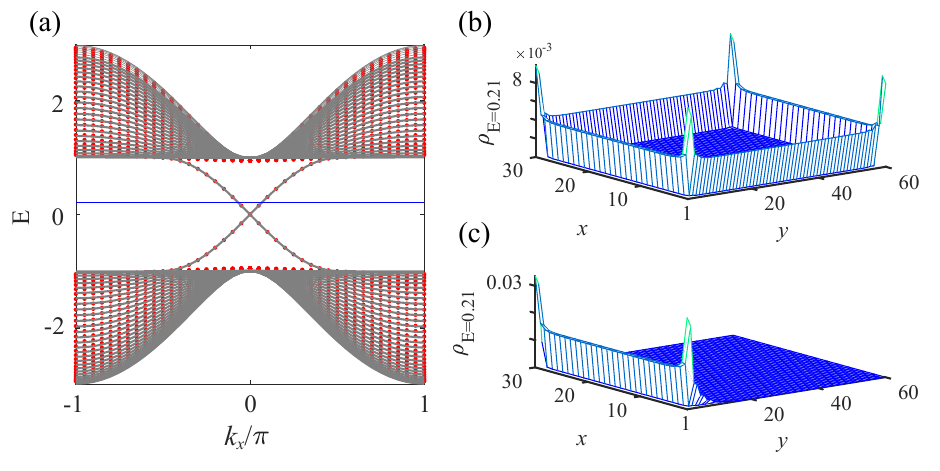}
\caption{(Color online) (a)Hermitian($\gamma=0$, grey) and non-Hermitian($\gamma=0.3$, red) Chern bands. (b) and (c) are the distributions of states with $E=0.21$ for Hermitian and non-Hermitian cases, respectively. The chiral edge states can be identified for the Hermitian case, but the distribution $\rho$ of topological edge modes losses its typical chirality for the non-Hermitian case due to the NHSE. We fix $M=-1$, $t=1$ and $N=30$.}
\label{f0}
\end{figure}

As the frontier of condensed matter physics and materials science, the study of topological phases have attracted widespread attention both theoretically and experimentally due to their fascinating properties
\cite{Chern1,Chern2,Chern3,Chern4,Chern5,Chern6,Chern7,Chern8,Chern9,Chern10,Chern11,Chern12,TI1,TI2,TI3,TI4,TI5,TI6,TI7,TI8,TI9,TI10,TI11,TI12,TI13,TI14,TI15,TI16,TI17,TI18,TI19,TI20,TI21,TSM1,TSM2,TSM3,TSM4,TSM5,TSM6,TSM7,TSM11,TSM12,TSM13,TSM14,TSM15,TSM16,TSM17,TSM18,TSM19,TSM20,TSM21,TSM22,TSM23,TSM24,TSM25,HOT1,HOT2,HOT3,HOT4,HOT5,HOT6,HOT7,HOT8}. In order to distinguish different types of topological phases, the topological invariants are introduced, and the phases with distinct topological invariants show different features \cite{BBC}. Typically, a Chern insulator with nonzero Chern number \cite{Chern4} will give rise to the edge modes located inside its nontrivial band gap [see Fig. \ref{f0} (a)]. The Chern number is linked to the number of the topological protected edge modes, which is important for practical applications. Thus, accurately identifying the Chern number of a sytem is fundamental for unlocking a wide range of functional capabilities \cite{Niu}. One illustrative instance is the domain wall \cite{TDW1,TDW2,TDW3,TDW4,TDW5,TDW6,TDW7,TDW8,TDW9,TDW10,TDW11,TDW12,TDW13,TDW14}, where the topological bound states emerge at the domain wall composed of two distinct Chern insulator phases. These features can be adopted to guide the design of low dissipation devices \cite{Device1,Device2,Device3,Device4,Device5,Device6,Device7,Device8,Device9}. Thus, it is crucial issue to measure the Chern number in experiments.

Generally, there are three schemes to measure the Chern number in experiments. Firstly, the quantized Hall conductance \cite{Chern1,Chern6,Chern12,TI4} is corresponding to Chern insulators and can be adopted to identify the Chern number. Secondly, due to the edge states emerging at the boundary [as shown in Fig. \ref{f0} (b)], the direct observation of the localized density distribution for edge states can reflect the nonzero Chern number of such systems \cite{Edge1,Edge2}.
Thirdly, the chirality of the edge modes, which corresponds to the sign of the Chern number, can be determined by observing the propagation direction of the current under an excitation at the boundary \cite{Chrirality1,Chrirality2}.

However, the above schemes only work for the Chern insulator in Hermitian cases.
Recently, the non-Hermitian topological systems \cite{NH1,NH2,NH3,NH4,NH5,NH6,NH7,NH8,NH9,NH12,NH13,Dis1,Dis2,Dis3,Dis4,Dis5,Mag1,Mag2,Dis6,Dis7,NHChern1,NHChern2,NHChern3,NHCI1,NHCI2,NHCI3}, especially non-Hermitian Chern insulators, have attracted significant interests due to their unique features.
For the non-Hermitian Hamiltonians, the energy spectrum under open boundary condition could be complex numbers, whose non-negligible imaginary part reflects considerable energy dissipation or gain \cite{NH1,NH2,NH9,NH10,NH11}.
Therefore, the validity and the quantization of the above transport properties are critically challenged in non-Hermitian Chern insulators \cite{NH4,Dis4,NHChern1,NHChern2,NHChern3}.
This implies that the corresponding Hall conductance fails to characterize the Chern number.
The study in Ref. \cite{NH3} further unveils that all eigenstates could be localized at the specific boundary, i.e. the non-Hermitian skin effect (NHSE).
These features leave the determination of the localized density distribution as well as the chirality for edge states [see Fig. \ref{f0} (c)] in non-Hermitian Chern insulators ambiguous.
Therefore, it is critically important to establish a clear experimental approach for determining the Chern number in non-Hermitian Chern insulators.

In this paper, we study the measurement of the topological invariants in non-Hermitian systems based on the magnetic effect. Recently, the study of Chern inulators under the magnetic field in Hermitian systems has demonstrated that the splitting bands relate to the Chern number \cite{HighChern}. We find that such discovery can get rid of the limitations of both local edge states and global transport characteristics. More importantly, this can dexterously overcome the drawbacks and identify the Chern number in non-Hermitian systems.
Motivated by this idea, we first study the influences of the magnetic field on the non-Hermitian Chern insulator.  When the non-Hermiticity is introduced, the open (closed) orbital sub-bands of the Hermitian system could evolve into the closed (open) orbital.
Furthermore, the non-Hermiticity can drive the phase transition from the Chern insulator to the normal insulator.
These results clearly manifest the distinctions between the Hermitian and non-Hermitian systems.
Significantly, we find that the splitting of non-Hermitian bands under a uniform magnetic field varies with the corresponding Chern number for all cases. These features make the detection of Chern number in non-Hermitian systems feasible in experiments.
To demonstrate this point, we simulate the detectable splitting spectra under the magnetic field in LC topological electric circuit as an example, and thus its Chern number can be measured.

This paper is organized as follows: In Sec. \ref{S2}, we study the non-Hermitian Chern insulator under the magnetic field. We illuminate the measurement of non-Hermitian spectra in LC electric circuit in Sec. \ref{S3}. Finally, a brief summary is presented in Sec. \ref{S5}.

\section{non-Hermitian Chern Insulators under the magnetic field}\label{S2}
In this section, we study the effects of the magnetic field in the non-Hermitian Chern insulator. We illuminate the magnetic effect can be utilized to identify the Chern number of non-Hermitian Chern bands.

\subsection{Model and Method}

To clarify the effects of the non-Hermiticity, we focus on the Chern insulator with the NHSE and the corresponding Hamiltonian reads \cite{NH4,Dis4,Dis7}:
\begin{align}
\begin{split}
\mathcal{H}=&\sum_{x,y}(M+2)\sigma_zc^{\dagger}_{x,y}c_{x,y}\\
&-t\frac{\sigma_z+i\sigma_x}{2}c^{\dagger}_{x,y}c_{x+1,y}-t\frac{\sigma_z-i\sigma_x}{2}c^{\dagger}_{x,y}c_{x-1,y}\\
&-t_+\frac{\sigma_z+i\sigma_y}{2}c^{\dagger}_{x,y}c_{x,y+1}-t_-\frac{\sigma_z-i\sigma_y}{2}c^{\dagger}_{x,y}c_{x,y-1},
\label{H}
\end{split}
\end{align}
where $M$ is the mass term.  $c^{\dagger}_{x,y}$ ($c_{x,y}$) represents the creation (annihilation) operator of site $(x,y)$. $\sigma_{x/y/z}$ are the Pauli matrices. The nonreciprocal hopping $t\pm\gamma$ is abbreviated as $t_{\pm}$. $t$ and $t_\pm$ are the hopping strength in $x$ and $y$ directions, respectively. Here, $\gamma$ denotes the strength of NHSE.
As illustrated by the density distribution in Fig. \ref{f0} (c), the typical chirality of the edge modes is destroyed when the NHSE exists ($\gamma\neq0$). We fix $t=1$ for simplicity.

We present the phase diagram to clarify the topological properties of the model. The momentum-space Hamiltonian $H(\textbf{k})$ satisfies $H(\textbf{k})|\psi^n_{R}(\textbf{k})\rangle=E_n(\textbf{k})|\psi^n_{R}(\textbf{k})\rangle, \langle \psi^n_{L}(\textbf{k})|H(\textbf{k})=E_n(\textbf{k})\langle \psi^n_{L}(\textbf{k})|$, and the normalization relation $\langle \psi^m_{L}(\textbf{k})|\psi^n_{R}(\textbf{k})\rangle=\delta_{mn}$. By considering $(k_x,k_y)\rightarrow(k_x,\tilde{k_y})$ with $\tilde{k_y}=k_y-i\ln(\beta)$ and $\beta=\sqrt{|\frac{t_-}{t_+}|}$, the non-Bloch Chern number \cite{NH4,Dis4} satisfies:
\begin{equation}
\textbf{C}=\frac{1}{2\pi i}\int d k_x d\tilde{k_y}Tr[P_{\tilde{\textbf{k}}} [\partial_{k_x}P_{\tilde{\textbf{k}}},\partial_{\tilde{k}_y}P_{\tilde{\textbf{k}}}]].
\label{C}
\end{equation}
Here, the projection operator is defined as $P_{\tilde{\textbf{k}}}=\sum_n|\psi^n_{R}(\tilde{\textbf{k}})\rangle \langle \psi^n_{L}(\tilde{\textbf{k}})|$. We emphasize that the Chern number $\textbf{C}$ is in bold face to distinguish it from the capacitor representation.

\begin{figure}[t]
\includegraphics[width=0.48\textwidth]{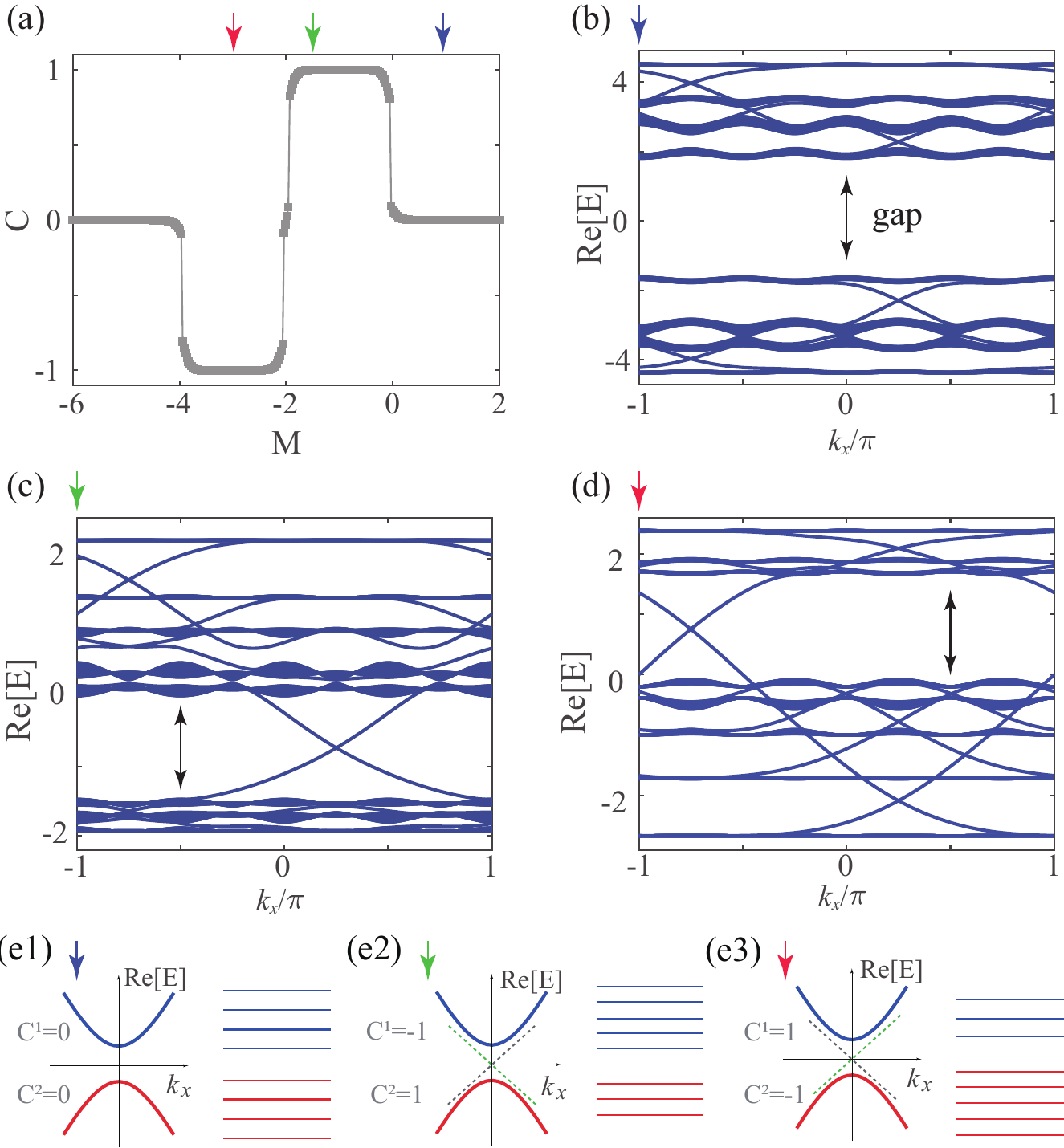}
\caption{(Color online) (a)The phase diagram of the non-Hermitian Chern insulator with non-Hermitian strength $\gamma=0.3$ and the sample size $N=20$. (b)-(d) are the plots of $Re[E]$ (the real part of the eigenvalues E) versus $k_x$ with $\frac{p}{q}=\frac{1}{4}$ and $\gamma=0.3$ for $M=1$, $M=-1.5$ and $M=-3$ corresponding to $\textbf{C}=0$, $\textbf{C}=1$ and $\textbf{C}=-1$, respectively. Here, the energy gap is marked by two-way arrows. (e1)-(e3) are schematic diagrams of the splitting sub-bands for (b)-(d), respectively. The Chern number for two parent bands ($\textbf{C}_{parent}$) are marked as $\textbf{C}^1$ and $\textbf{C}^2$. We fix $t=1$ for simplicity.}
\label{f1}
\end{figure}

In the Fig. \ref{f1} (a), the non-Bloch Chern number $\textbf{C}$ versus $M$ is obtained, which is consistent with the theoretical predicted phase boundaries $M=[\sqrt{t_+t_-}-1,-\sqrt{t_+t_-}-1,\sqrt{t_+t_-}-3,-\sqrt{t_+t_-}-3]$ [see the Appendix \ref{A} for more details]. For $\textbf{C}\neq0$, the edge modes in the bulk energy gap can be identified, shown in Figs. \ref{f0}(a) and (b). Unfortunately, the distribution of edge states losses its typical chirality due to the existence of the NHSE, as shown in Fig. \ref{f0}(c). Thus, the Chern number is difficult to be determined experimentally compared with the Hermitian cases.

\subsection{The Chern number's measurement based on the splitting sub-bands}

Next, we investigate the magnetic effect in the non-Hermitian Chern insulator based on the above model.

We consider a magnetic field $\textbf{B}=(0,0,B)$ along z direction with the magnetic flux per unit cell $\Phi=BS=\frac{p}{q}\Phi_0$. The unit cell area $S=a^2=1$ and $\Phi_0=h/e$ is the quantum flux. $p$ and $q$ give the number of splits of Landau levels and non-Bloch bands, respectively \cite{Niu2}. The Hamiltonian in real space can be rewritten as:
\begin{align}
\begin{split}
\mathcal{H}_{B}=&\sum_{x,y}(M+2)\sigma_zc^{\dagger}_{x,y}c_{x,y}\\
&-te^{i2\pi\frac{p}{q}y}\frac{\sigma_z+i\sigma_x}{2}c^{\dagger}_{x,y}c_{x+1,y}\\
&-te^{-i2\pi\frac{p}{q}y}\frac{\sigma_z-i\sigma_x}{2}c^{\dagger}_{x,y}c_{x-1,y}\\
&-t_+\frac{\sigma_z+i\sigma_y}{2}c^{\dagger}_{x,y}c_{x,y+1}-t_-\frac{\sigma_z-i\sigma_y}{2}c^{\dagger}_{x,y}c_{x,y-1}.
\label{HB}
\end{split}
\end{align}
 Notably, we find that the splits of the energy spectrum can be adopted to distinguish its topology.
To clarify the Chern number dependence of the splitting sub-bands in non-Hermitian Chern insulators, we focus on the following three typical cases with $M=1$, $M=-1.5$ and $M=-3$. Their Chern numbers are $\textbf{C}=0$, $\textbf{C}=1$ and $\textbf{C}=-1$, as marked in Fig. \ref{f1} (a) (arrows). In the following, the non-Hermitian terms are fixed at $\gamma=0.3$ unless otherwise stated.

As shown in Figs. \ref{f1} (b)-(d), we plot $Re[E]$ versus $k_x$ under the magnetic field with $\frac{\Phi}{\Phi_0}=\frac{p}{q}=\frac{1}{4}$. Due to the existence of the magnetic effect, two parent bands (the initial bands with $\frac{\Phi}{\Phi_0}=0$) split into $2q=8$ sub-bands. Such a feature depends only on $q$, and independent for the topology of the system. Nevertheless, we notice that the band splitting shows distinct distributions for different $\textbf{C}$, which is also reported in the corresponding Hermitian system \cite{HighChern}. Notably, the sub-bands are symmetrically distributed with respect to the energy gap for $\textbf{C}=0$ [see Fig. \ref{f1} (b)], while the sub-bands are asymmetrically distributed for $\textbf{C}\neq0$. For example, there are five sub-bands above the initial bulk energy gap and three below the gap for $\textbf{C}=1$ in Fig. \ref{f1} (c). Conversely, for $\textbf{C}=-1$, the distribution is reversed, with three sub-bands above and five below the gap, as shown in Fig. \ref{f1} (d). Note that the chirality between $\textbf{C}=1$ and $\textbf{C}=-1$ is opposite, and the splitting of the energy spectrum with the higher Chern number is also different \cite{HighChern}. Thus, the splitting of bands shows distinct features for systems with distinct Chern number $\textbf{C}$. As illustrated schematically in Figs. \ref{f1} (e1)-(e3), the red and blue bands provide a clearer visualization.

The picture can be understood by the semicalssical theory \cite{Niu1,Niu2,Semi}. The external magnetic field $\textbf{B}=(0,0,B)$ can be decomposed into two terms: $B=B_0+\delta B$. Here, $B_0=0$ relates to the non-Bloch bands and $\delta B=B$ represents a perturbation by the magnetic field.
The effective Lagrangian is
\begin{equation}
\mathcal{L}(\textbf{r},\tilde{\textbf{k}},\dot{\textbf{r}},\dot{\tilde{\textbf{k}}})=-e\delta \textbf{A}(\textbf{r})\cdot \dot{\textbf{r}}+\hslash\tilde{\textbf{k}}\cdot\dot{\textbf{r}}+\hslash\mathcal{A}(\tilde{\textbf{k}})\cdot\dot{\tilde{\textbf{k}}}-E(\tilde{\textbf{k}})
\end{equation}
where $E(\tilde{\textbf{k}})=\mathcal{E}(\tilde{\textbf{k}})+e/2m\delta\textbf{B}\cdot\textbf{L}(\tilde{\textbf{k}})$ [$\mathcal{E}(\tilde{\textbf{k}})$ is the reduced form of the non-Bloch band $E(\tilde{\textbf{k}})$] and $\boldsymbol{\Omega}(\tilde{\textbf{k}})=\nabla\times\mathcal{A}(\tilde{\textbf{k}})$. Here, $\textbf{L}(\tilde{\textbf{k}})$ and $\boldsymbol{\Omega}(\tilde{\textbf{k}})$ are the magnetic moment and Berry curvature, respectively. The gauge for $\delta \textbf{B}$ is $\delta \textbf{A}(\textbf{r})=\frac{1}{2}\delta \textbf{B}\times\textbf{r}$. Note that $\tilde{\textbf{k}}$ ranges over the generalized Brillouin zone and the magnetic field should be $\delta \textbf{B}$.
By considering the relationship $\dot{\textbf{r}}=\dot{\tilde{\textbf{k}}}\times \widehat z (\hslash/e\delta B)$, the above equation in the momentum $\tilde{\textbf{k}}$ reads:
\begin{equation}
\mathcal{L}(\tilde{\textbf{k}},\dot{\tilde{\textbf{k}}})=\frac{\hslash^2}{2e\delta B}\tilde{\textbf{k}}\cdot(\dot{\tilde{\textbf{k}}}\times \widehat z) +\hslash\mathcal{A}(\tilde{\textbf{k}})\cdot\dot{\tilde{\textbf{k}}}-E(\tilde{\textbf{k}}).
\end{equation}
The generalized momentum of $\tilde{\textbf{k}}$ is
\begin{equation}
\boldsymbol{\pi}=\frac{\partial \mathcal{L}}{\partial \dot{\tilde{\textbf{k}}}}=-\frac{\hslash^2}{2e\delta B}\tilde{\textbf{k}}\times \widehat z +\hslash\mathcal{A}(\tilde{\textbf{k}}).
\end{equation}
Following the Bohr-Sommerfeld quantization condition, the quantization rule for ``hyperorbit" $\mathcal{C}_m$ is given by
\begin{align}
\begin{split}
\oint_{\mathcal{C}_m}\boldsymbol{\pi}\cdot d\tilde{\textbf{k}}&=\oint_{\mathcal{C}_m}[-\frac{\hslash^2}{2e\delta B}\tilde{\textbf{k}}\times \widehat z +\hslash\mathcal{A}(\tilde{\textbf{k}})]\cdot d\tilde{\textbf{k}}\\
&=\frac{\hslash^2}{2e\delta B}\oint_{\mathcal{C}_m}(\tilde{\textbf{k}}\times d\tilde{\textbf{k}})\cdot\widehat z+\hslash\oint_{\mathcal{C}_m}\mathcal{A}(\tilde{\textbf{k}})d\tilde{\textbf{k}}\\
&=(m+\frac{1}{2})h.
\end{split}
\end{align}
Thus, according to the restraint on the outermost orbit $\mathcal{C}_{max}$, the area of $\mathcal{C}_{max}$ should satisfy
\begin{align}
\begin{split}
&\frac{1}{2q}\oint_{\mathcal{C}_{max}}(\tilde{\textbf{k}}\times d\tilde{\textbf{k}})\cdot\widehat z\\
\geq&\frac{(2\pi/a)^2}{q}[m+\frac{1}{2}-\frac{1}{2\pi}\oint_{\mathcal{C}_{max}}\mathcal{A}(\tilde{\textbf{k}})d\tilde{\textbf{k}}]\delta \phi
\label{BSQ}
\end{split}
\end{align}
where $\frac{1}{2\pi}\oint_{\mathcal{C}_{max}}\mathcal{A}(\tilde{\textbf{k}})d\tilde{\textbf{k}}\approx-\textbf{C}_{parent}$ (``$-$" comes from the fact that the integral trajectory is in the opposite direction) and $m=0,1,2\cdots$. $\textbf{C}_{parent}$ is the Chern number of a parent band.

\begin{table}[t]
\setlength\tabcolsep{4.5pt}
\centering
\caption{The Chern number of parent bands and sub-bands for $\frac{p}{q}=\frac{1}{4}$. These sub-bands are numbered from maximum to minmum of the real eigenvalue $Re[E]$.}
\label{table1}
\begin{tabular}{c|c|c|c|c|c|c|c|c|c|c} \hline % 其中，|c|表示文本居中，文本两边有竖直表线。
$\gamma$ & M & $\textbf{C}_{parent}$ & $\textbf{C}_{1}$ &$\textbf{C}_{2}$ &$\textbf{C}_{3}$&$\textbf{C}_{4}$&$\textbf{C}_{5}$&$\textbf{C}_{6}$&$\textbf{C}_{7}$&$\textbf{C}_{8}$  \\ \hline\hline
0 & 1 & (0;0) & -1 &-1 & 3&-1&-1&3&-1&-1  \\ \hline
0 & -1.5 & (-1;1) & -1 &-1 & 3&{\color{red}\textbf{3}}&{\color{red}\textbf{-5}}&3&-1&-1  \\ \hline
0 & -2.5 & (1;-1) & -1 &-1 & 3&{\color{red}\textbf{-5}}&{\color{red}\textbf{3}}&3&-1&-1  \\ \hline
0 & -3 & (1;-1) & -1 &-1 & 3&\textbf{3}&\textbf{-1}&-1&-1&-1  \\ \hline
0 & -3.75 & \textbf{(1;-1)} & -1 &-1 & 3&-1&-1&3&-1&-1  \\ \hline\hline
0.3 & 1 & (0;0) & -1 &-1 & 3&-1&-1&3&-1&-1  \\ \hline
0.3 & -1.5 & (-1;1)&  -1 &-1 & 3&\textbf{-1}&\textbf{-1}&3&-1&-1 \\ \hline
0.3 & -2.5 & (1;-1)&  -1 &-1 & 3&\textbf{-1}&\textbf{-1}&3&-1&-1 \\ \hline
0.3 & -3 & (1;-1) & -1 &-1 & 3&\textbf{3}&\textbf{-1}&-1&-1&-1  \\ \hline\hline
0.9 & -3 & (1;-1) & -1 &-1 & 3&{\color{red}\textbf{4}}&{\color{red}\textbf{-2}}&-1&-1&-1  \\ \hline
0.9 & -3.75 & {\color{red}\textbf{(0;0)}} & -1 &-1 & 3&-1&-1&3&-1&-1  \\ \hline\hline
\end{tabular}
\label{T1}
\end{table}

Based on Eq. (\ref{BSQ}), one has
\begin{align}
\begin{split}
m\leq\frac{1}{\delta \phi}-\frac{1}{2}-\textbf{C}_{parent}.
\end{split}
\end{align}
This quantization rule determines the number of sub-bands under the magnetic field $\delta \phi=1/q$:
\begin{equation}
N_s=|q-\textbf{C}_{parent}|.
\label{Key}
\end{equation}
According to Eq. (\ref{Key}), the parent band with $\textbf{C}_{parent}=1,0,-1$ will split into $N_s=3,4,5$ sub-bands for $p/q=1/4$ respectively, which is consistent with the numerical results in Fig. \ref{f1}.

From an experimental standpoint, measuring the topology of a non-Hermitian system is of utmost importance.
Our above studies imply that whether the Chern number of a non-Hermitian energy spectrum is nonzero can be judged by changing the direction of the magnetic field.
Specifically, when the magnetic field is reversed, the energy spectrum of a trivial system does not change at all, while a non-trivial system changes because of $N_s=|-q-\textbf{C}_{parent}|=|q+\textbf{C}_{parent}|$. More importantly, the chirality of non-Hermitian Chern insulators can also be identified by reversing the magnetic field. The `$\pm$' chirality of the Chern number can be verified by the sign of $\textbf{C}_{parent}=(|q+\textbf{C}_{parent}|-|q-\textbf{C}_{parent}|)/2$.
Such measurement schemes only depend on the splitting of the energy spectrum and still work in non-Hermitian systems.
Contrarily, conventional characteristic methods \cite{Chern1,Chern6,Chern12,TI4,Edge1,Edge2,Chrirality1,Chrirality2}, e.g. the measurement of Hall conductance, the distribution as well as the direction of propagation for edge states, fail.
Thus, it can be used as a strong indicator for experimentally detecting the Chern number in non-Hermitian Chern insulators by counting the number of sub-bands $N_s$ under the magnetic field. Note that there have been several related experiments investigating the magnetic effect in non-Hermitian systems \cite{MagExperient1,MagExperient2}.% 加文献

\subsection{Differences between Hermitian and non-Hermitian Chern bands}
\begin{figure}[t]
\includegraphics[width=0.48\textwidth]{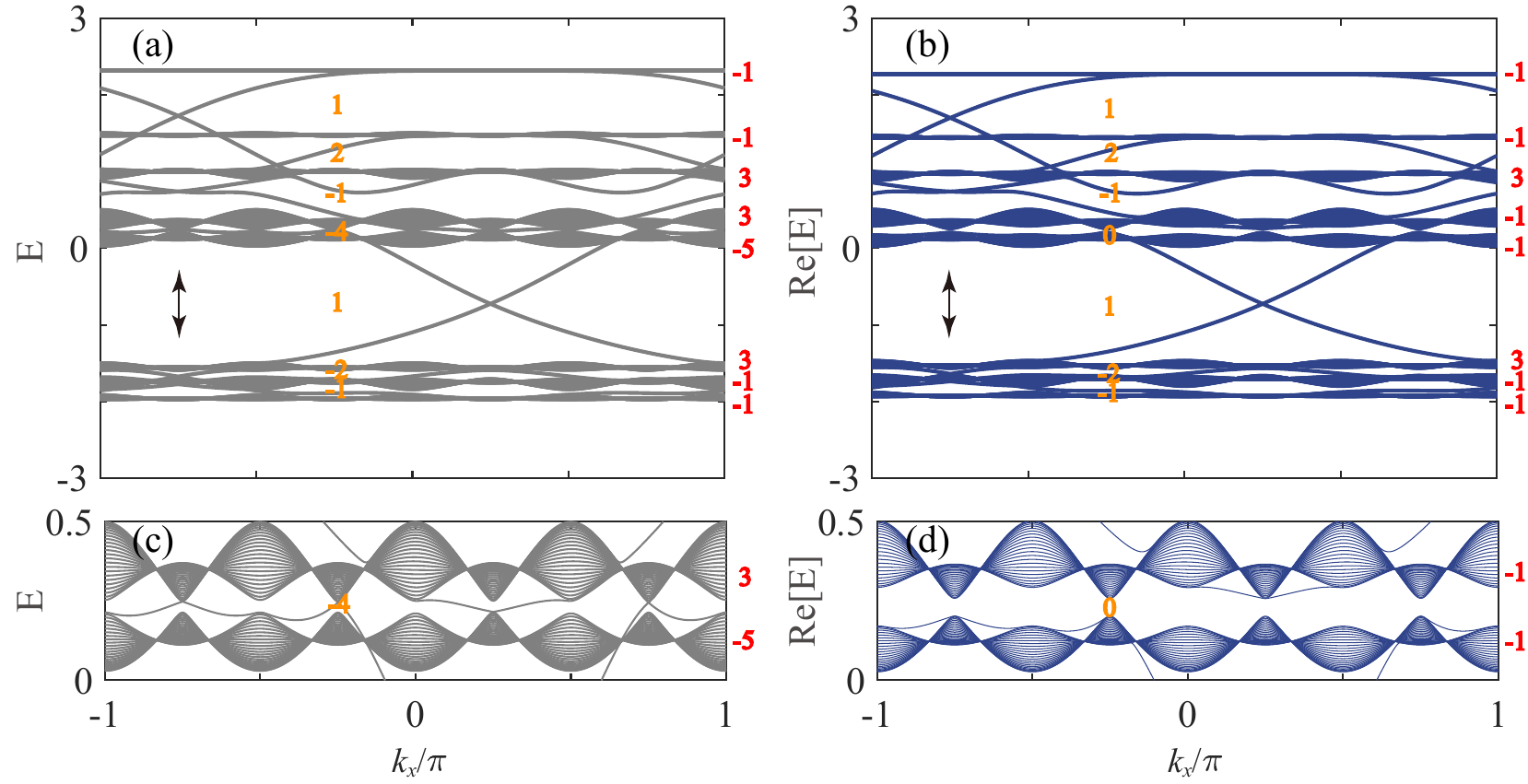}
\caption{(Color online) (a)Hermitian ($\gamma=0$) and (b) Non-Hermitian ($\gamma=0.3$) energy spectra for $\frac{p}{q}=\frac{1}{4}$, and $M=-1.5$. Orange numbers are the number of the edge states of each energy gap and red numbers are the Chern number $C_m$ of sub-bands for the Chern band, where $\pm$ describes the chirality. (c)-(d) are the zoom of (a)-(b) for the middle two sub-bands with the Chern number $\textbf{C}_q$ and $\textbf{C}_{q+1}$, respectively.}
\label{f2}
\end{figure}

The splitting band technique effectively addresses the challenges in measuring the Chern number that arise from the non-Hermiticity. It also plays a pivotal role in detecting unique phenomena associated with non-Hermitian properties. Here, we show the novel phenomena induced by the magnetic field in non-Hermitian Chern insulators, and check the power of the proposed method.

Firstly, the energy spectrum for non-Hermitian systems shows distinctive features compared with their Hermitian counterparts. To elucidate the distinction, we calculate the Chern number of each sub-band $\textbf{C}_m$.
The Berry curvature of each sub-band reads
\begin{align}
\begin{split}
\boldsymbol{\Omega}_m(\tilde{\textbf{k}})=i\sum_{n\neq m}&[\frac{\langle \psi^m_{L}|\partial_{k_x}h|\psi^n_{R}\rangle \langle \psi^n_{L}|\partial_{\tilde{k_y}}h|\psi^m_{R}\rangle}{(E_n-E_m)^2}\\
&-\frac{\langle \psi^m_{L}|\partial_{\tilde{k_y}}h|\psi^n_{R}\rangle \langle \psi^n_{L}|\partial_{k_x}h|\psi^m_{R}\rangle}{(E_n-E_m)^2}].
\end{split}
\end{align}
Here $m\in[1,2q]$ because the Chern insulator is a two-band model. Note that the Hamiltonian $h$ is a $2q\times2q$ matrix:
$$
h
=
\left[
	\begin{matrix}
 D_1& T_{y1}  & \textbf{0} & \cdots   & T_{y2}   \\
T_{y2} & D_2  & T_{y1}& \cdots   & \textbf{0}  \\
\vdots  & \vdots  & \ddots & \vdots  & \vdots  \\
\textbf{0}   & \cdots  & T_{y2} & D_{q-1} & T_{y1}\\
T_{y1}   & \cdots  & \textbf{0} & T_{y2} & D_q  \\
\end{matrix}
	\right]
$$
with $D_k=[(M+2)-\cos(k_x+2\pi (k-1)\frac{p}{q})]\sigma_z+\sin(k_x+2\pi (k-1)\frac{p}{q})\sigma_x,T_{y1}=t_+\beta\frac{-\sigma_z-i\sigma_y}{2}e^{ik_y}$ and $T_{y2}=t_-/\beta\frac{-\sigma_z+i\sigma_y}{2}e^{-ik_y}.$
The Chern number of each sub-band is obtained:
\begin{align}
\begin{split}
\textbf{C}_m=\frac{1}{2\pi}\int d\tilde{\textbf{k}}\boldsymbol{\Omega}_m(\tilde{\textbf{k}}).
\label{Cm}
\end{split}
\end{align}
For clearness, the corresponding $\textbf{C}_m$ is listed in TAB. \ref{T1} for different $\gamma$ and $M$ with $p/q=1/4$ (Here, $\gamma=0$ is adopted from the Hermitian case in \cite{HighChern}).

As shown in Fig. \ref{f1} and corroborated by the previous study \cite{Mag2}, the splitting of Chern bands for non-Hermitian systems follows the same principles observed in Hermitian systems [see Fig. \ref{A1} in the Appendix \ref{C}].
Nevertheless, we notice that the Chern number of the sub-bands ($\textbf{C}_q$ and $\textbf{C}_{q+1}$) could be distinct from the Hermitian cases. For the Hermitian case $M=-1.5$, the open-orbit sub-bands exhibit $\textbf{C}_q=q-1$ and $\textbf{C}_{q+1}=-q-1$ [see red numbers in Fig. \ref{f2} (a)] \cite{HighChern}. However, the Chern numbers show the closed-orbit features for $\gamma=0.3$ with $\textbf{C}_q=-1$ and $\textbf{C}_{q+1}=-1$, as shown in Fig. \ref{f2} (b).
An enlarged view of relevant details is presented in Figs. \ref{f2}(c) and (d).
Even though $\textbf{C}_{parent}$ of the two parent bands are the same for both cases, a trivial (non-trivial) energy gap without (with) edge modes can be identified for $\gamma=0.3$ ($\gamma=0$).
Furthermore, the second-order splitting bands will be remarkably different for both cases due to variation of the Chern number for splitting sub-bands (e.g. $\textbf{C}_{4}=3$ and $\textbf{C}_{5}=-5$ for the Hermitian case, and $\textbf{C}_{4/5}=-1$ for the non-Hermitian) \cite{Niu1,Niu2}.
Actually, these features can be experimentally observed by the proposed method based on Eq. (\ref{Key}).

The reasons of the evolution of orbits are related to the ratio between the hopping intensities along the $x$ direction ($t_x=t$) and $y$ direction ($t_y=t_\pm$) \cite{Hof1,Hof2,Hof3}.
Based on the non-Bloch band theory \cite{NH4,NH3}, the effective hopping in $y$ direction satisfies the following relation:
\begin{equation}
t_y\rightarrow \tilde{t}_y=\sqrt{t_+t_-}=\sqrt{t^2-\gamma^2}.
\end{equation}
When $t_y=t$ changes into $\tilde{t}_y=\sqrt{t_+t_-}$, $t_x/\tilde{t}_y$ can continuously change by varying $\gamma$ \cite{Hof1,Hof2,Hof3}. Then, the open-orbit sub-bands with $\textbf{C}_q=q-1$ and $\textbf{C}_{q+1}=-q-1$ could turn into the closed orbits with $\textbf{C}_q=-1$ and $\textbf{C}_{q+1}=-1$. Such exotic feature in non-Hermitian Chern insulators is not unique for $M=-1.5$ and $p/q=1/4$. It is also available for other $M$ and $p/q$ cases, e.g. $M=-2.5$ and $p/q=1/5$ [see TABs. \ref{T1} and \ref{T2}].

\begin{figure}[t]
\includegraphics[width=0.48\textwidth]{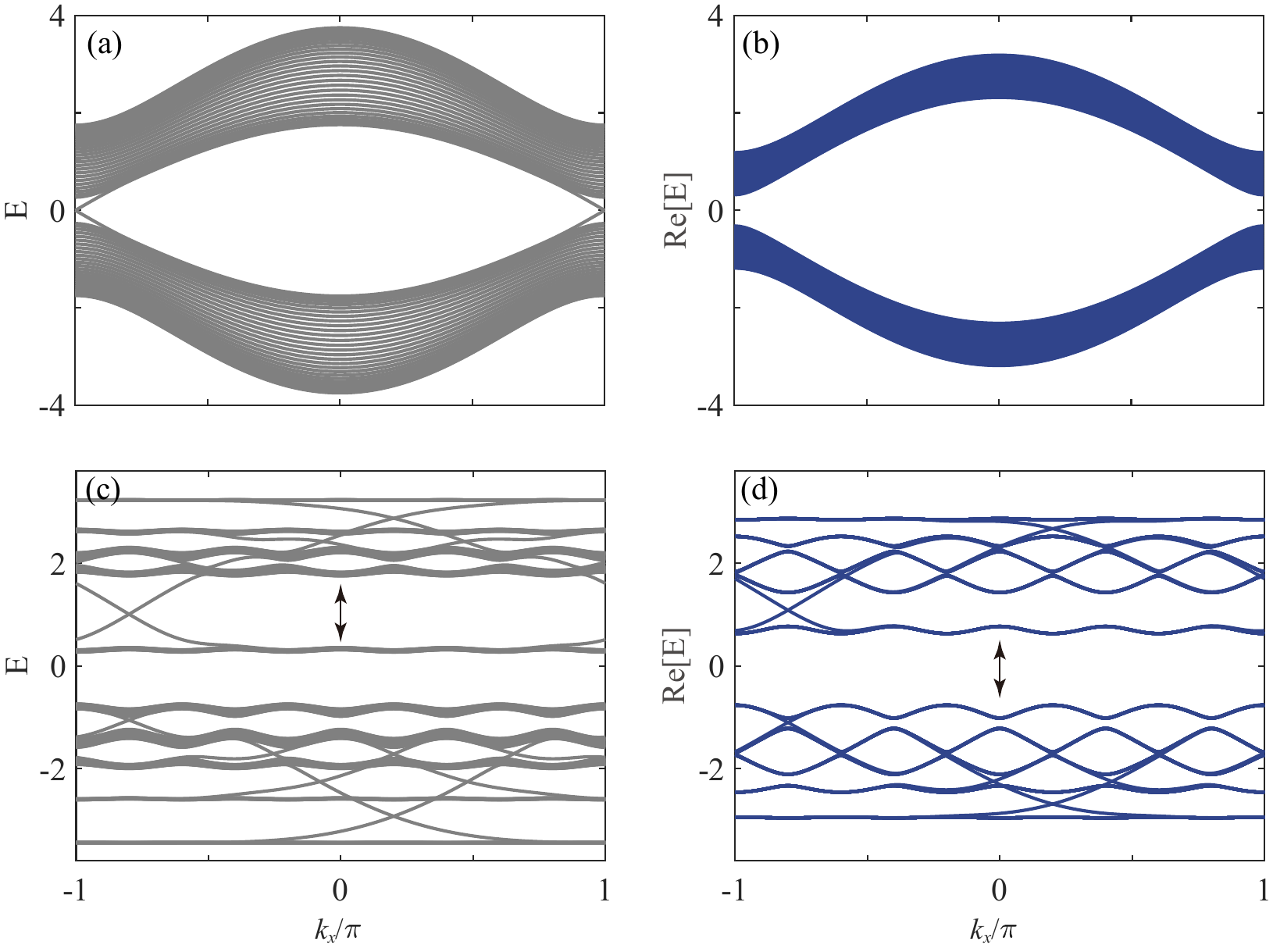}
\caption{(Color online) (a)Hermitian ($\gamma=0$) and (b) Non-Hermitian ($\gamma=0.9$) energy spectra for $M=-3.75$. (c)-(d) are the splitting bands of (a)-(b) when the magnetic field is considered, for example $\frac{p}{q}=\frac{1}{5}$.}
\label{f3}
\end{figure}

Secondly, the non-Hermiticity can drive the topological phase transition from the Chern insulator phase to the normal insulator phase, which is also reported in the previous studies \cite{NH4,Dis4}. We study the splitting bands for these two cases in Fig. \ref{f3}. In the Hermitian limits [see Figs. \ref{f3} (a) and (c)], the edge states exist in the band gap because of the Chern number $\textbf{C}=-1$. The two parent bands split into $2q=10$ sub-bands under the magnetic field $\frac{p}{q}=\frac{1}{5}$, with 6 sub-bands located below the bulk energy gap. The strong non-Hermiticity can induce a phase transition from the Chern insulator phase with $\textbf{C}=-1$ to the normal insulator phase with $\textbf{C}=0$, as shown in Figs. \ref{f3} (b) and (d). For the normal insulator phase, the edge modes disappear and its two parent bands become trivial, i.e. $\textbf{C}_{parent}=0$. When the magnetic field is applied, the 10 sub-bands are symmetrically distributed on both sides of the energy gap, which satisfies with the theory $N_s=|q-\textbf{C}_{parent}|=5$. The complete phase diagram is shown in the Appendix \ref{A}.
The proposed method enables the observation of distinct splitting sub-bands between Hermitian and non-Hermitian systems. In consequence, this technique is highly effective in identifying phase transitions in non-Hermitian Chern insulators. It is important to note that, although theories regarding these phase transitions have been previously proposed \cite{NH4,Dis4}, their experimental verification has not yet been achieved.

\section{Measurement of the non-Hermitian band in LC Circuit}\label{S3}

As uncovered in the above section, topological characteristics of non-Hermitian Chern insulators can be measurable in principle by utilizing the magnetic field. It successfully addresses the challenges associated with conventional detection methods. The key of this detection method is the measurement and analysis of the splitting bands. Subsequently, we demonstrate the implement of measurement method in LC electric circuit since the design and fabrication are relatively easy.

\begin{figure}[t]
\includegraphics[width=0.48\textwidth]{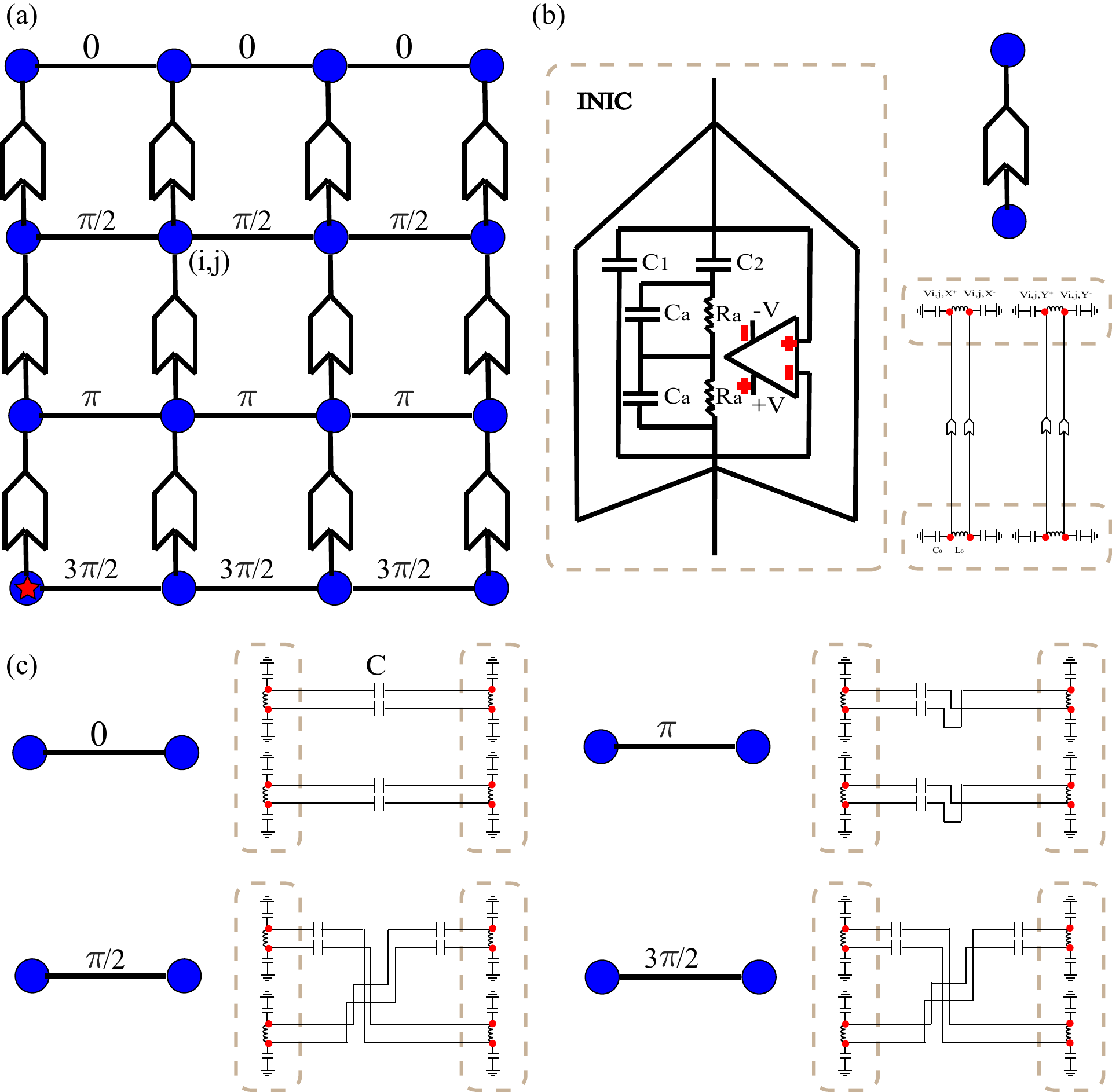}
\caption{(Color online) (a)The circuit design for a one-band non-Hermitian system with the NHSE in y direction under the magnetic field with $\frac{p}{q}=\frac{1}{4}$. Ac voltage source is placed at the downer-leftmost corner site marked by a red star. (b) and (c) are the hoppings in y and x directions, respectively. The negative impedance converters with current inversion (INIC) introduces the nonreciprocal nearest-neighbor hopping $C_1\pm C_2$.}
\label{f4}
\end{figure}

Without loss of generality, we consider a square lattice with one-band Hamiltonian (the trivial two-dimensional electronic gas) under the uniform magnetic field with $\frac{\Phi}{\Phi_0}=\frac{1}{4}$:
\begin{align}
\begin{split}
\mathcal{H}^1_{B}=&\sum_{x,y}te^{i\frac{\pi}{2}y}c^{\dagger}_{x,y}c_{x+1,y}+te^{-i\frac{\pi}{2}y}c^{\dagger}_{x,y}c_{x-1,y}\\
&+t_+c^{\dagger}_{x,y}c_{x,y-1}+t_-c^{\dagger}_{x,y}c_{x,y+1}.
\label{HB1}
\end{split}
\end{align}
Note that the NHSE exists along the $y$ direction if $\gamma\neq0$.
The realization of the model is illustrated in Fig. \ref{f4}. Each site $(i,j)$ (blue dots) is composed of two inductors $L_0$ with four nodes (red dots). These four nodes' voltages are marked as  $[V_{i,j,X^+}, V_{i,j,X^-},V_{i,j,Y^+}, V_{i,j,Y^-}]$, so the voltages of two $L_0$ satisfy  $U_{i,j,X}=V_{i,j,X^+}-V_{i,j,X^-}$ and $U_{i,j,Y}=V_{i,j,Y^+}-V_{i,j,Y^-}$. The voltage of the site $(i,j)$ is $U_{i,j}=U_{i,j,X}+iU_{i,j,Y}$ \cite{LC1,LC2}. In addition, the grounded capacitors $C_0$ are connected to $L_0$ to stabilize the circuit. To introduce the NHSE, the negative impedance converters with current inversion (INIC) \cite{INIC1,INIC2,INIC3,INIC4} is adopted to obtain the nonreciprocal nearest-neighbor hopping. As shown in Fig. \ref{f4} (b), two distinct hopping strengths $C_1\pm C_2$ are available along the $y$ direction (We fix $C_1>C_2$ and $C_1=C$ for simplicity). Here, DC power suppliers provide $\pm10$V DC voltage for Op-Amps (LT1363). Moreover, the nearest-neighbor hopping $t$ in the transverse ($x$) direction is realized with the help of a capacitor $C$. Finally, the corresponding magnetic flux is obtained by braiding the capacitive couplings [see Fig. \ref{f4} (c)] \cite{Flux,Flux1}. An AC voltage source is placed at the lower-left-corner site marked by a red star only for measurement.
Compared to other experimental systems, LC circuits take one great advantage: the voltage corresponds to the amplitude of the wave function.
It can be measured easily, and non-Hermitian bands will be obtained based on discrete Fourier transform of the voltage \cite{DFT}.

\begin{figure}[b]
\includegraphics[width=0.48\textwidth]{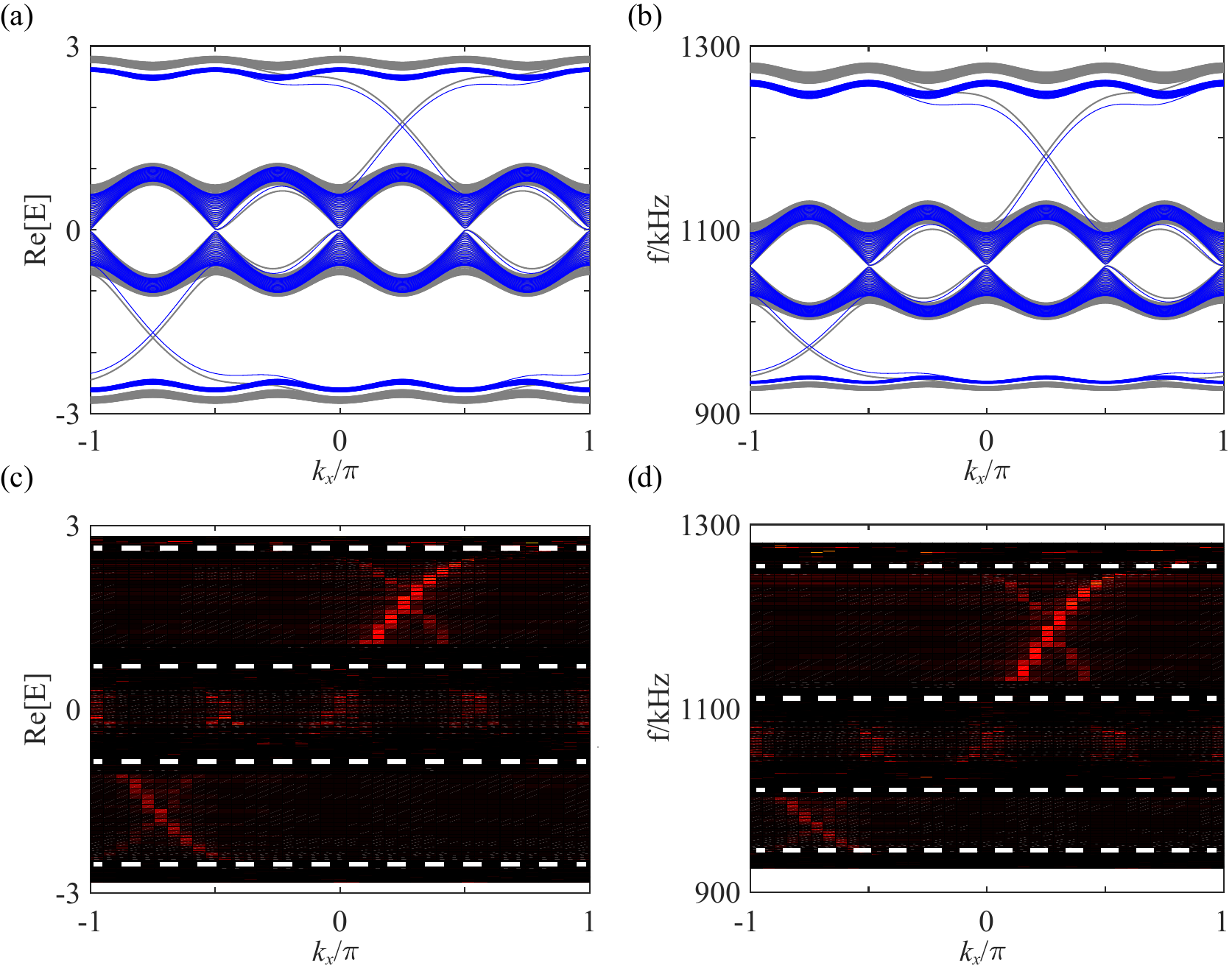}
\caption{(Color online) The energy spectra and frequency spectra versus $k_x$ for the one-band model with Eq. (\ref{HB1}). (a) and (b) The infinite-size samples with Hermitian (grey,$\gamma=0$) and non-Hermitian (blue,$\gamma=0.5$). (c) and (d)The finite-size samples ($N_x=40$ and $N_y=20$) for $\gamma=0.05$. White dashed lines represent the splitting sub-bands.  We fix $L_0=5.0uH$, $C_0=5nF$ and $C=1nF$.}
\label{f5}
\end{figure}
\begin{figure}[t]
\includegraphics[width=0.48\textwidth]{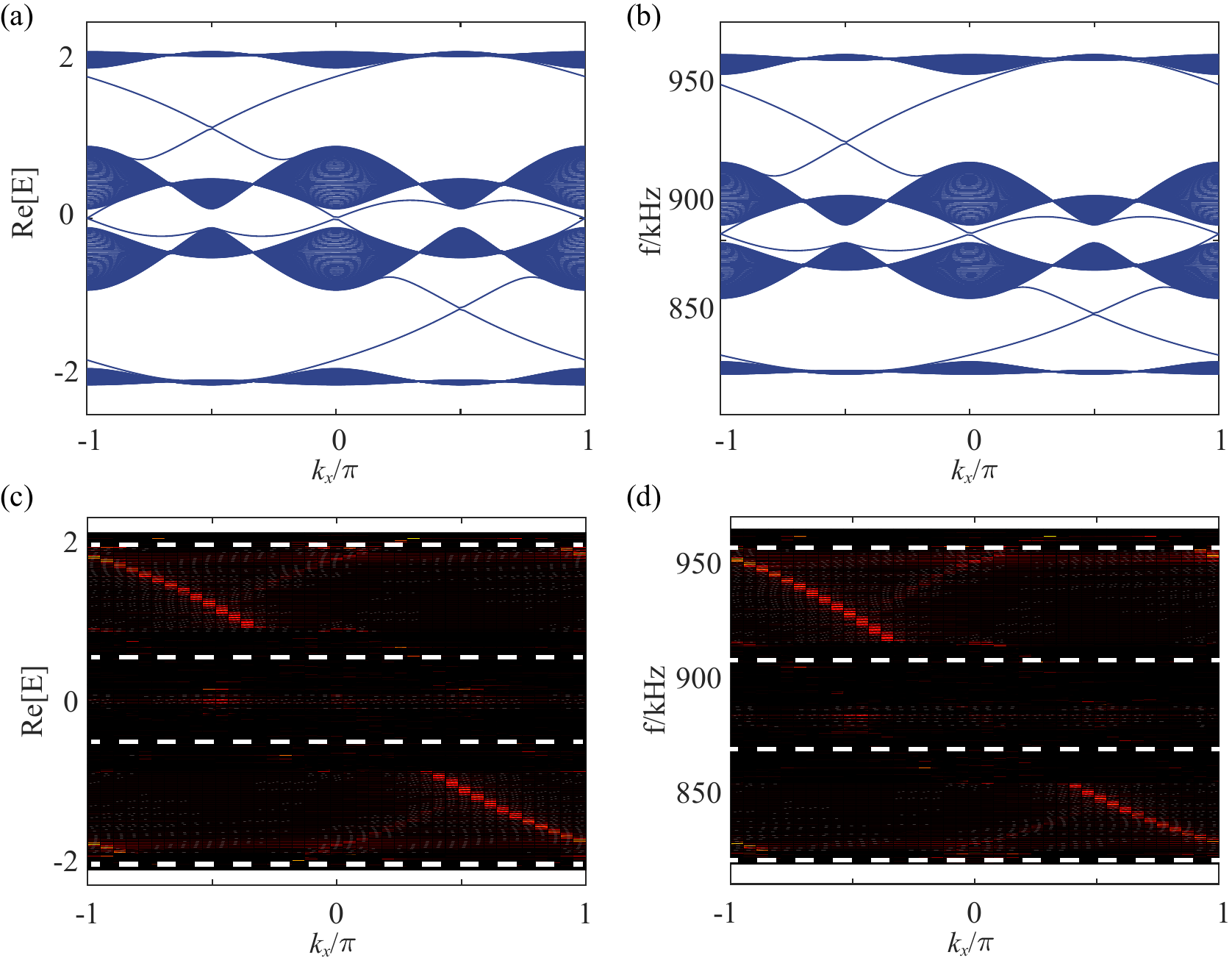}
\caption{(Color online) The energy spectra and frequency spectra versus $k_x$ the non-Hermitian Chern insulator Eq. (\ref{HB}) with $M=-1.5$ and $p/q=1/2$. (a) and (b) The infinite-size samples with the non-Hermitian strength $\gamma=0.05$. (c) and (d) are same as (a)and (b), except the finite-size samples ($N_x=40$ and $N_y=20$).  White dashed lines represent the splitting sub-bands.  We fix $L_0=5.0uH$, $C_0=5nF$ and $C=1nF$.}
\label{f6}
\end{figure}

\begin{figure}[b]
\includegraphics[width=0.48\textwidth]{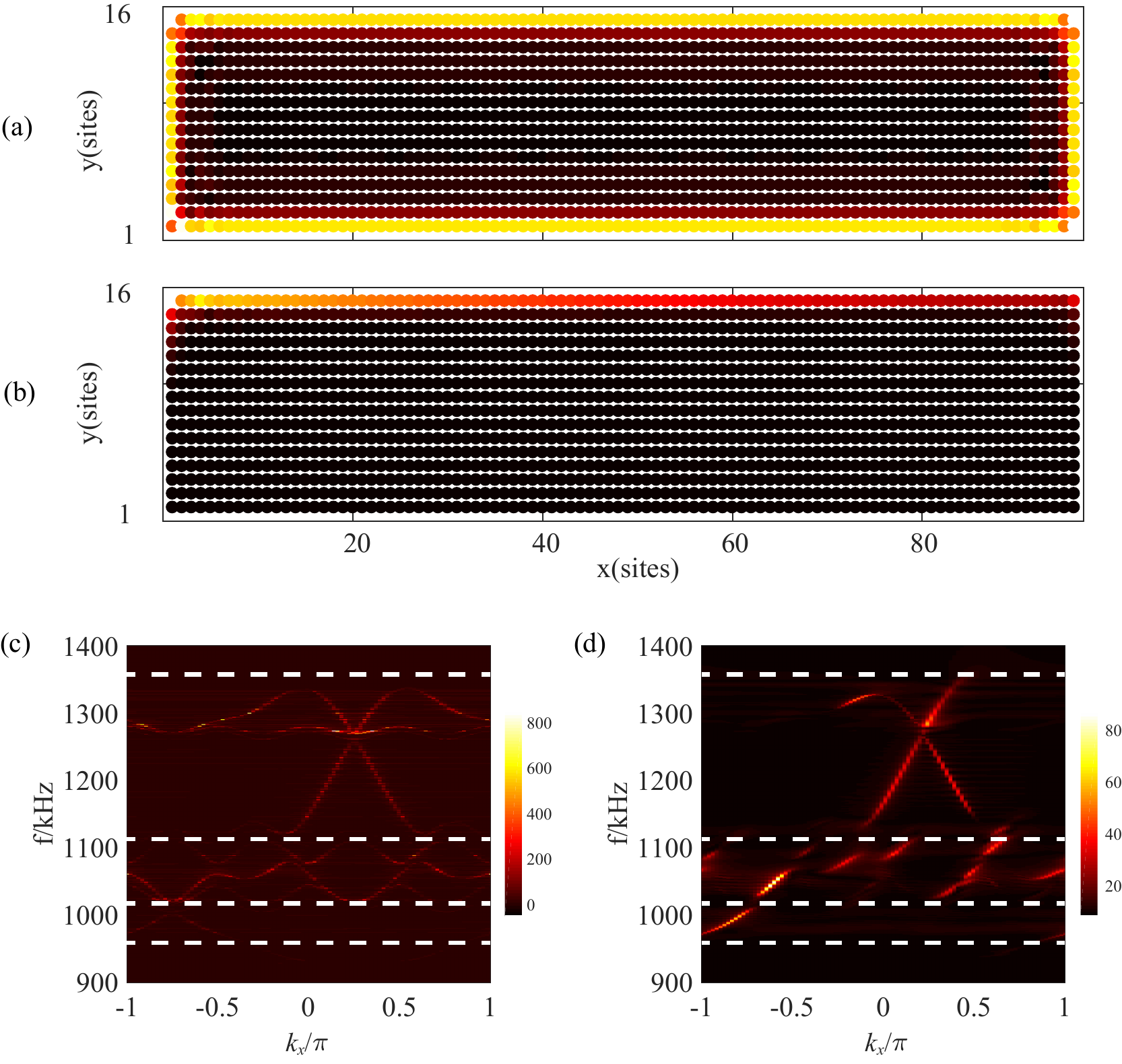}
\caption{(Color online) (a) and (b) are the distribution of voltages of $f=1200.4kHz$ for Hermitian ($C_1=1nF,C_2=0pF$) and non-Hermitian ($C_1=1nF,C_2=500pF$), respectively. (c) and (d) are $f$ versus $k_x$ based on discrete Fourier transform for Hermitian ($C_1=1nF,C_2=0pF$) and non-Hermitian ($C_1=1nF,C_2=100pF$) in LC electric circuit. White dashed lines represent the splitting sub-bands. We fix $L_0=5.0uH$, $C_0=5nF$, $C=1nF$, $R_a=20\Omega$, $C_a=1nF$, $+V=10V$, $-V=-10V$, $N_x=96$ and $N_y=16$.}
\label{f7}
\end{figure}

Next, we give a numerical calculation of the corresponding spectra by discrete Fourier transform. According to the Kirchhoff's laws [see the Appendix \ref{B} for more details], the relationship of frequency and eigenvalue for the model Eq. (\ref{HB1}) is:
\begin{align}
\begin{split}
f=\frac{1}{2\pi}\frac{1}{\sqrt{L_0C}}\sqrt{\frac{2}{C_0/C+4-E}}.
\label{eqB6}
\end{split}
\end{align}
Here, $\omega=2\pi f$ and $E$ is the eigenvalue of the corresponding Hamiltonian.
In this circuit network setup, the nonreciprocal hopping with $C_1+C_2$ and $C_1-C_2$ induces the NHSE in the $y$ direction.

In Figures \ref{f5} (a) and (b), we plot the energy and frequency spectra versus $k_x$ for both the Hermitian (grey,$\gamma=0$) and non-Hermitian (blue,$\gamma=0.5$) cases of Eq. (\ref{HB1}) and Eq. (\ref{eqB6}) for the infinite-size samples. Since $\frac{p}{q}=\frac{1}{4}$, the band splits into $q=4$ sub-bands. According to Eq. (\ref{Key}), the splitting sub-bands imply the trivial feature of the system ($\textbf{C}=\textbf{C}_{parent}=0$) when the magnetic field disappears, i.e. $N_s=|q-\textbf{C}_{parent}|=|4-0|=4$.

In practical experimental contexts, constructing a large-scale platform poses significant challenges. Thus, we also study the finite-size samples. For the sample with size $N_x=40$ and $N_y=20$, the wave function (voltage) is calculated numerically. Based on discrete Fourier transform for the wave function (voltage), the corresponding non-Hermitian spectra [see Figs. \ref{f5} (c) and (d)] can be obtained. Four splitting sub-bands are observed, which is consistent with Figs. \ref{f5} (a) and (b). Here, the white dashed lines represent the splitting sub-bands.

We also consider non-Hermitian Chern insulators [Eq. (\ref{HB})] with $\textbf{C}=1$ under the magnetic field with $\frac{\Phi}{\Phi_0}=\frac{p}{q}=\frac{1}{2}$.
The realization of this model is very similar to Fig. \ref{f4}, except by considering two sublattices A and B with $[V^A_{i,j,X^+}, V^A_{i,j,X^-},V^A_{i,j,Y^+}, V^A_{i,j,Y^-}]$ and $[V^B_{i,j,X^+}, V^B_{i,j,X^-},V^B_{i,j,Y^+}, V^B_{i,j,Y^-}]$ [see the Appendix \ref{B} for more details]. The magnetic flux is obtained by only braiding the capacitive coupling with $0$ and $\pi$ in Fig. \ref{f4} (c). In this case, the frequency reads:
\begin{align}
\begin{split}
f=\frac{1}{2\pi}\frac{1}{\sqrt{L_0C}}\sqrt{\frac{2}{C_0/C+8-E}}
\label{eqB7}
\end{split}
\end{align}
In Figs. \ref{f6} (a) and (b), we plot the energy and frequency spectra versus $k_x$ for the model with Eq. (\ref{HB}). Since $\frac{p}{q}=\frac{1}{2}$, the band splits into $q=4$ sub-bands. One sub-band below the gap is observed. According to Eq. (\ref{Key}), $N_s=|q-\textbf{C}_{parent}|=|2-1|=1$, which suggests the Chern number $\textbf{C}=1$.
For the sample with size $N_x=40$ and $N_y=20$, the spectra obtained by discrete Fourier transform [see Figs. \ref{f6} (c) and (d)] are consistent with (a) and (b).
These results demonstrate the feasibility of our proposal.

In order to strengthen the proposed method, we further simulate the topological electric circuit with $N_x=96$ and $N_y=16$ for the model Eq. (\ref{HB1}). All circuit simulations are performed based on the LTspice (www.linear.com/LTspice). Firstly, the localized density distributions of the edge states for the non-Hermitian cases are influenced by the NHSE, and are totally distinct from the Hermitian ones for $f=1200.4kHz$ [see Figs. \ref{f7}(a) and (b)]. In consequence, the edge modes for the non-Hermitian systems are difficult to be directly identified in experiments. However, our proposal overcomes these difficulties. To demonstrate it, the frequency spectra are obtained based on discrete Fourier transform by sweeping frequency [see Figs. \ref{f7}(c) and (d)]. The band splits into four sub-bands, which are consistent with the theoretical results. Based on the distribution of these sub-bands, we can faithfully measure the Chern number of the systems experimentally.

In short, the splitting frequency spectra in the magnetic field can be obtained through frequency sweep methods. Following this, the topology of non-Hermitian Chern bands can be inferred. This method greatly simplifies the complex task of directly examining the spatial distribution of topological edge states or the transport properties of systems that are significantly affected by the NHSE.
Thus, the Chern number of the non-Hermitian Chern insulator can be obtained by the splitting of the energy band under the magnetic field, which varies with topology of the systems.

\section{summary}\label{S5}

In summary, we studied the interplay of the non-Hermitian Chern insulator and the magnetic effect.
We revealed that the magnetic field can be utilized to detect the topological invariant of the non-Hermitian systems. The number of splitting sub-bands are closely related to the Chern number of each parent band, which can be utilized to determine the Chern number of non-Hermitian Chern insulators.
Experimentally, the measurement of the Chern number under the magnetic field is illustrated in the LC topological electric circuit by sweeping frequency.
Our work addresses the challenge of measurement for topological properties in non-Hermition systems in experiments.

\section{ACKNOWLEDGEMENT}
We are grateful for fruitful discussions with Qiang Wei, Mian Peng, Jing Yu and Haiwen Liu.
This work is financially supported by the National Basic Research Program of China (Grants No. 2019YFA0308403 and No. 2022YFA1403700), National Natural Science Foundation of China under Grants No. 12350401, No. 12204044 and No. 12147126, and a Project Funded by the Priority Academic Program Development of Jiangsu Higher Education Institutions.

\appendix

\section{The phase diagram of the non-Hermitian Chern insulator}\label{A}

\begin{figure}[b]
\includegraphics[width=0.48\textwidth]{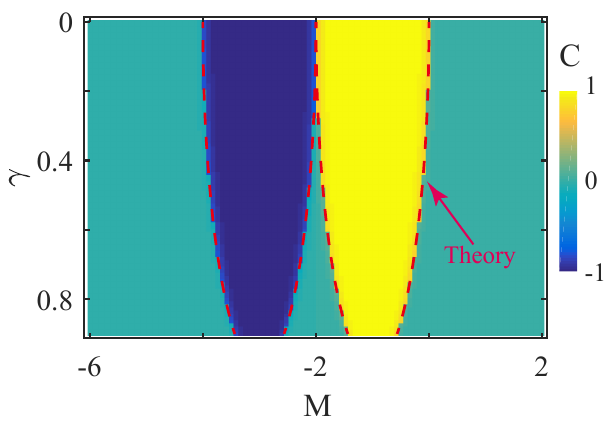}
\caption{(Color online) The phase diagram of the non-Hermitian Chern insulator. The red solid line is obtained based on the theoretical analysis of the phase transition $M=[\sqrt{t_+t_-}-1,-\sqrt{t_+t_-}-1,\sqrt{t_+t_-}-3,-\sqrt{t_+t_-}-3]$. We fix sample size $N=20$.}
\label{A0}
\end{figure}

In the Appendix, we show the phase diagram of the non-Hermitian Chern insulator, as shown in Fig. \ref{A0}. Based on the non-Bloch theory, the Hamiltonian $\mathcal{H}(k_x,\tilde{k_y})$ reads:
\begin{align}
\begin{split}
\mathcal{H}(k_x,\tilde{k_y})=&(M+2)\sigma_z\\
-&t\frac{\sigma_z+i\sigma_x}{2}exp(-ik_x)-t\frac{\sigma_z-i\sigma_x}{2}\exp(ik_x)\\
-&\sqrt{t_+t_-}\frac{\sigma_z+i\sigma_y}{2}\exp(-ik_y)\\
-&\sqrt{t_+t_-}\frac{\sigma_z-i\sigma_y}{2}\exp(ik_y)\\
=&d_x\sigma_x+d_y\sigma_y+d_z\sigma_z
\end{split}
\end{align}
with $d_x=-t\sin(k_x), d_y=-\sqrt{t_+t_-}\sin(k_y)$ and $d_z=(M+2)-\sqrt{t_+t_-}\cos(k_y)-t\cos(k_x)$. Here, $k_x\in [0,2\pi]$ and $k_y\in [0,2\pi]$. When $d_z=0$, the gapless Dirac cone exists at $(k_x,k_y)=[(0,0);(0,\pi);(\pi,0);(\pi,\pi)]$, which gives the phase transition points $M=[\sqrt{t_+t_-}-1,-\sqrt{t_+t_-}-1,\sqrt{t_+t_-}-3,-\sqrt{t_+t_-}-3]$ (red dashed lines).
The numerical results are consistent with the theoretical predictions. Note that the non-Hermiticity can induce the phase transition from the Chern insulator phase with $\textbf{C}\neq0$ to the normal insulator phase $\textbf{C}=0$, which is also reported in the previous studies \cite{NH4,Dis4}.

\begin{figure}[b]
\includegraphics[width=0.48\textwidth]{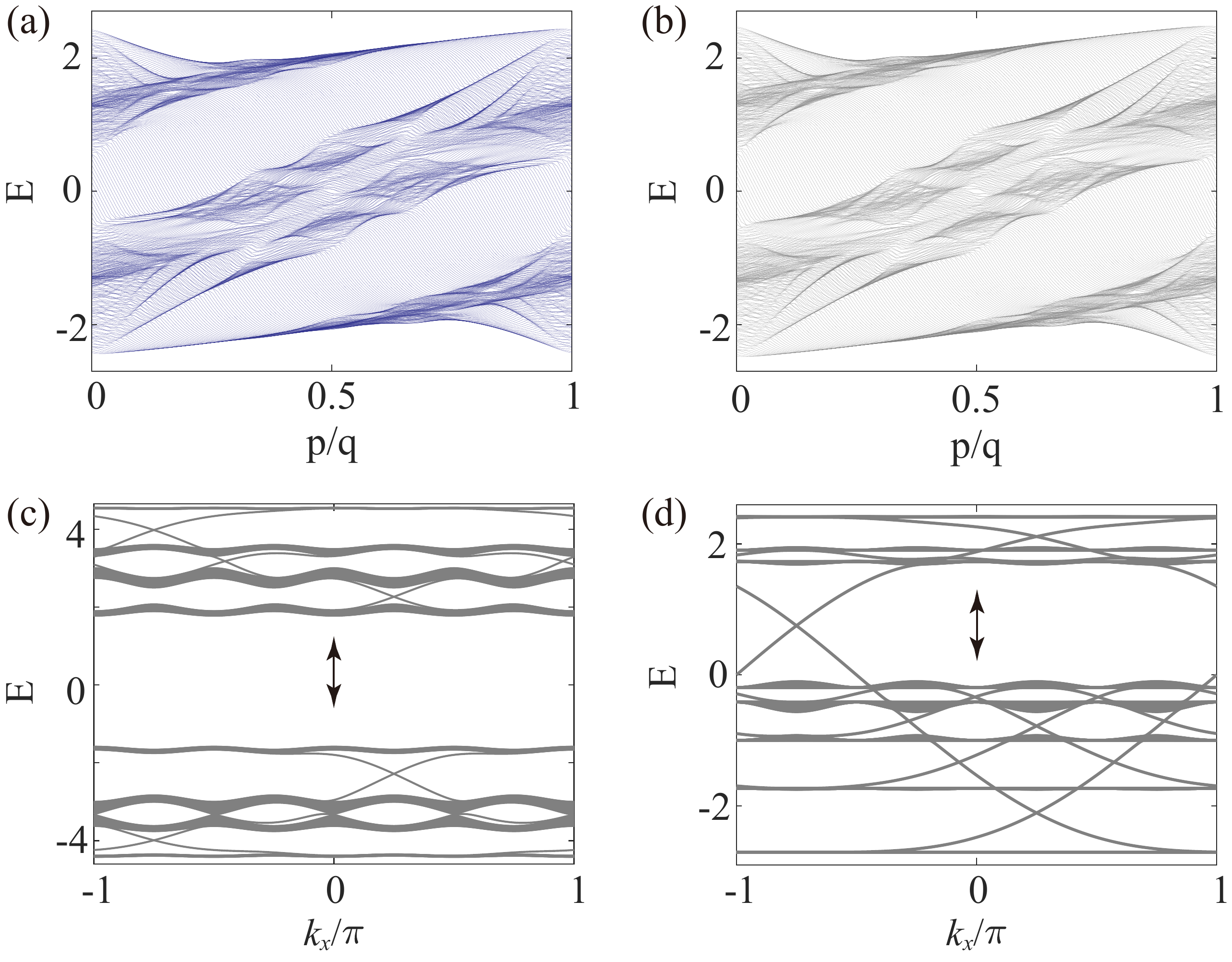}
\caption{(Color online) (a)(b)The Hofstadter butterfly of the Chern insulator for $\gamma=0.3$ (non-Hermitian) and $\gamma=0$ (Hermitian) with $M=-1.5$. (c) and (d) are the plots of $E$ versus $k_x$ with $\frac{p}{q}=\frac{1}{4}$ and $\gamma=0$ (grey) for $M=1$ and $M=-3$ corresponding to $\textbf{C}=0$ and $\textbf{C}=-1$, respectively. We fix $t=1$ for simplicity.}
\label{A1}
\end{figure}

\section{The splitting sub-bands and the Chern number}\label{C}

Compared with Fig. \ref{f1}, Fig. \ref{A1} and the previous study \cite{Mag2}, the splitting of Chern band for non-Hermitian cases seems to be the same as the Hermitian under the magnetic field. However, we find the unique non-Hermitian features by analyzing the Chern number $\textbf{C}_m$ of the sub-bands, as shown in Fig. \ref{f2} and TAB. \ref{T1} in the main text, as well as TAB. \ref{T2} (Here, $\gamma=0$ is the Hermitian case \cite{HighChern}).
Specifically, the closed (open)-orbit sub-bands can also evolve into the open (closed)-orbit by increasing the non-Hermitian strength $\gamma$.

\begin{table}[t]
\setlength\tabcolsep{2.5pt}
\centering
\caption{The Chern number of parent bands and sub-bands for $\frac{p}{q}=\frac{1}{5}$. These sub-bands are numbered from maximum to minmum of the real eigenvalue $Re[E]$.}
\label{table2}
\begin{tabular}{c|c|c|c|c|c|c|c|c|c|c|c|c} \hline % 其中，|c|表示文本居中，文本两边有竖直表线。
$\gamma$ & M & $\textbf{C}_{parent}$ & $\textbf{C}_{1}$ &$\textbf{C}_{2}$ &$\textbf{C}_{3}$&$\textbf{C}_{4}$&$\textbf{C}_{5}$&$\textbf{C}_{6}$&$\textbf{C}_{7}$&$\textbf{C}_{8}$&$\textbf{C}_{9}$&$\textbf{C}_{10}$  \\ \hline\hline
0 & 1 & (0;0) & -1 &-1 & 4&-1&-1&-1&4&-1&-1&-1  \\ \hline
0 & -1.5 & (-1;1) & -1 &4 & -6&4&{\color{red}\textbf{4}}&{\color{red}\textbf{-6}}&4&-1&-1&-1  \\ \hline
0 & -2.5 & (1;-1) & -1 &-1 & -1&4&{\color{red}\textbf{-6}}&{\color{red}\textbf{4}}&4&-6&4&-1  \\ \hline
0 & -3 & (1;-1) & -1 &-1 & -1&4&\textbf{4}&\textbf{-1}&\textbf{-1}&-1&-1&-1  \\ \hline
0 & -3.75 & \textbf{(1;-1)} & -1 &-1 & -1&4&-1&-1&-1&4&-1&-1  \\ \hline\hline
0.3 & 1 & (0;0) & -1 &-1 & 4&-1&-1&-1&4&-1&-1&-1  \\ \hline
0.3 & -1.5 & (-1;1)& -1 &4 & -6&4&\textbf{-1}&\textbf{-1}&4&-1&-1&-1 \\ \hline
0.3 & -2.5 & (1;-1) & -1 &-1 & -1&4&\textbf{-1}&\textbf{-1}&4&-6&4&-1  \\ \hline
0.3 & -3 & (1;-1) & -1 &-1 & -1&4&\textbf{4}&\textbf{-1}&\textbf{-1}&-1&-1&-1  \\ \hline\hline
0.9 & -3 & (1;-1) & -1 &-1 & -1&4&{\color{red}\textbf{5}}&{\color{red}\textbf{-7}}&{\color{red}\textbf{4}}&-1&-1&-1  \\ \hline
0.9 & -3.75 & {\color{red}\textbf{(0;0)}} & -1 &-1 & -1&4&-1&-1&-1&4&-1&-1  \\ \hline\hline
\end{tabular}
\label{T2}
\end{table}
\section{Derivation of Eq. (\ref{eqB6}) and Eq. (\ref{eqB7})}\label{B}
Firstly, we show the derivation of Eq. (\ref{eqB6}). According to the Kirchhoff's laws, the circuit equations of nodes $X^+, X^-, Y^+, Y^-$ of $L_0$ at the site $(i,j)$ read:
\begin{align}
\begin{split}
I_{i,j,X^+}&=\frac{V_{i,j,X^+}-V_{i,j,X^-}}{i\omega L_0}+i\omega C_0 V_{i,j,X^+}\\
&+i\omega C[(V_{i,j,X^+}-V_{i-1,j,Y^+})+(V_{i,j,X^+}-V_{i+1,j,Y^-})]\\
&+i\omega (C_1+C_2)(V_{i,j,X^+}-V_{i,j-1,X^+})\\
&+i\omega (C_1-C_2)(V_{i,j,X^+}-V_{i,j+1,X^+})
\end{split}
\end{align}
\begin{align}
\begin{split}
I_{i,j,X^-}&=\frac{V_{i,j,X^-}-V_{i,j,X^+}}{i\omega L_0}+i\omega C_0 V_{i,j,X^-}\\
&+i\omega C[(V_{i,j,X^-}-V_{i-1,j,Y^-})+(V_{i,j,X^-}-V_{i+1,j,Y^+})]\\
&+i\omega (C_1+C_2)(V_{i,j,X^-}-V_{i,j-1,X^-})\\
&+i\omega (C_1-C_2)(V_{i,j,X^-}-V_{i,j+1,X^-})
\end{split}
\end{align}
\begin{align}
\begin{split}
I_{i,j,Y^+}&=\frac{V_{i,j,Y^+}-V_{i,j,Y^-}}{i\omega L_0}+i\omega C_0 V_{i,j,Y^+}\\
&+i\omega C[(V_{i,j,Y^+}-V_{i-1,j,X^-})+(V_{i,j,Y^+}-V_{i+1,j,X^+})]\\
&+i\omega (C_1+C_2)(V_{i,j,Y^+}-V_{i,j-1,Y^+})\\
&+i\omega (C_1-C_2)(V_{i,j,Y^+}-V_{i,j+1,Y^+}),
\end{split}
\end{align}
and
\begin{align}
\begin{split}
I_{i,j,Y^-}&=\frac{V_{i,j,Y^-}-V_{i,j,Y^+}}{i\omega L_0}+i\omega C_0 V_{i,j,Y^-}\\
&+i\omega C[(V_{i,j,Y^-}-V_{i-1,j,X^+})+(V_{i,j,Y^-}-V_{i+1,j,X^-})]\\
&+i\omega (C_1+C_2)(V_{i,j,Y^-}-V_{i,j-1,Y^-})\\
&+i\omega (C_1-C_2)(V_{i,j,Y^-}-V_{i,j+1,Y^-}).
\end{split}
\end{align}
Straightforward, we have
\begin{align}
\begin{split}
U_{i,j,X}=&-\frac{i\omega L_0}{2}[(i\omega C_0+4i\omega C)U_{i,j,X}\\
&+i\omega C(-U_{i-1,j,Y}+U_{i+1,j,Y})\\
&-i\omega (C_1+C_2)U_{i,j-1,X}-i\omega (C_1-C_2)U_{i,j+1,X}]
\end{split}
\end{align}
and
\begin{align}
\begin{split}
U_{i,j,Y}=&-\frac{i\omega L_0}{2}[(i\omega C_0+4i\omega C)U_{i,j,Y}\\
&+i\omega C(U_{i-1,j,X}-U_{i+1,j,X})\\
&-i\omega (C_1+C_2)U_{i,j-1,Y}-i\omega (C_1-C_2)U_{i,j+1,Y}].
\end{split}
\end{align}
Note that we consider $C_1=C$ only for simplicity.
The voltage of the site $(i,j)$ is
\begin{align}
\begin{split}
U_{i,j}=&U_{i,j,X}+iU_{i,j,Y}\\
=&(\frac{\omega^2 L_0 C_0}{2}+2\omega^2 L_0 C)U_{i,j}+\frac{i\omega^2 L_0 C}{2}(U_{i-1,j}-U_{i+1,j})\\
&-\frac{\omega^2 L_0}{2}(C_1+C_2)U_{i,j-1}-\frac{\omega^2 L_0}{2}(C_1-C_2)U_{i,j+1},
\end{split}
\end{align}
and then the coefficient of $U_{i,j}$ gives \cite{LC1,LC2}
\begin{align}
\begin{split}
E=\frac{C_0}{C}+4-\frac{2}{\omega^2 L_0 C}.
\end{split}
\end{align}
Therefore,
\begin{align}
\begin{split}
f=\frac{1}{2\pi}\frac{1}{\sqrt{L_0C}}\sqrt{\frac{2}{C_0/C+4-E}}.
\end{split}
\end{align}
Here, $\omega=2\pi f$.

For the two-band Chern insulator [Eq. (\ref{H}), here we fix $C_1=C$ for simplicity], the circuit equations of nodes $X^+, X^-, Y^+, Y^-$ of $L_0$ at the site $(i,j)$ for the sublattices A and B are:
\begin{align}
\begin{split}
&I^A_{i,j,X^+}\\
=&\frac{V^A_{i,j,X^+}-V^A_{i,j,X^-}}{i\omega L_0}+i\omega C_0 V^A_{i,j,X^+}\\
+&i\omega C[(V^A_{i,j,X^+}-V^A_{i-1,j,X^-})+(V^A_{i,j,X^+}-V^B_{i-1,j,Y^+})\\
+&(V^A_{i,j,X^+}-V^A_{i+1,j,X^-})+(V^A_{i,j,X^+}-V^B_{i+1,j,Y^-})]\\
+&i\omega (C_1+C_2)[(V^A_{i,j,X^+}-V^A_{i,j-1,X^-})+(V^A_{i,j,X^+}-V^B_{i,j-1,X^+})]\\
+&i\omega (C_1-C_2)[(V^A_{i,j,X^+}-V^A_{i,j+1,X^-})+(V^A_{i,j,X^+}-V^B_{i,j+1,X^-})],
\end{split}
\end{align}
\begin{align}
\begin{split}
&I^A_{i,j,X^-}\\
=&\frac{V^A_{i,j,X^-}-V^A_{i,j,X^+}}{i\omega L_0}+i\omega C_0 V^A_{i,j,X^-}\\
+&i\omega C[(V^A_{i,j,X^-}-V^A_{i-1,j,X^+})+(V^A_{i,j,X^-}-V^B_{i-1,j,Y^-})\\
+&(V^A_{i,j,X^-}-V^A_{i+1,j,X^+})+(V^A_{i,j,X^-}-V^B_{i+1,j,Y^+})]\\
+&i\omega (C_1+C_2)[(V^A_{i,j,X^-}-V^A_{i,j-1,X^+})+(V^A_{i,j,X^-}-V^B_{i,j-1,X^-})]\\
+&i\omega (C_1-C_2)[(V^A_{i,j,X^-}-V^A_{i,j+1,X^+})+(V^A_{i,j,X^-}-V^B_{i,j+1,X^+})],
\end{split}
\end{align}
\begin{align}
\begin{split}
&I^A_{i,j,Y^+}\\
=&\frac{V^A_{i,j,Y^+}-V^A_{i,j,Y^-}}{i\omega L_0}+i\omega C_0 V^A_{i,j,Y^+}\\
+&i\omega C[(V^A_{i,j,Y^+}-V^A_{i-1,j,Y^-})+(V^A_{i,j,Y^+}-V^B_{i-1,j,X^-})\\
+&(V^A_{i,j,Y^+}-V^A_{i+1,j,Y^-})+(V^A_{i,j,Y^+}-V^B_{i+1,j,X^+})]\\
+&i\omega (C_1+C_2)[(V^A_{i,j,Y^+}-V^A_{i,j-1,Y^-})+(V^A_{i,j,Y^+}-V^B_{i,j-1,Y^+})]\\
+&i\omega (C_1-C_2)[(V^A_{i,j,Y^+}-V^A_{i,j+1,Y^-})+(V^A_{i,j,Y^+}-V^B_{i,j+1,Y^-})],
\end{split}
\end{align}
and
\begin{align}
\begin{split}
&I^A_{i,j,Y^-}\\
=&\frac{V^A_{i,j,Y^-}-V^A_{i,j,Y^+}}{i\omega L_0}+i\omega C_0 V^A_{i,j,Y^-}\\
+&i\omega C[(V^A_{i,j,Y^-}-V^A_{i-1,j,Y^+})+(V^A_{i,j,Y^-}-V^B_{i-1,j,X^+})\\
+&(V^A_{i,j,Y^-}-V^A_{i+1,j,Y^+})+(V^A_{i,j,Y^-}-V^B_{i+1,j,X^-})]\\
+&i\omega (C_1+C_2)[(V^A_{i,j,Y^-}-V^A_{i,j-1,Y^+})+(V^A_{i,j,Y^-}-V^B_{i,j-1,Y^-})]\\
+&i\omega (C_1-C_2)[(V^A_{i,j,Y^-}-V^A_{i,j+1,Y^+})+(V^A_{i,j,Y^-}-V^B_{i,j+1,Y^+})];
\end{split}
\end{align}
\begin{align}
\begin{split}
&I^B_{i,j,X^+}\\
=&\frac{V^B_{i,j,X^+}-V^B_{i,j,X^-}}{i\omega L_0}+i\omega C_0 V^B_{i,j,X^+}\\
+&i\omega C[(V^B_{i,j,X^+}-V^A_{i-1,j,Y^+})+(V^B_{i,j,X^+}-V^B_{i-1,j,X^+})\\
+&(V^B_{i,j,X^+}-V^A_{i+1,j,Y^-})+(V^B_{i,j,X^+}-V^B_{i+1,j,X^+})]\\
+&i\omega (C_1+C_2)[(V^B_{i,j,X^+}-V^A_{i,j-1,X^-})+(V^B_{i,j,X^+}-V^B_{i,j-1,X^+})]\\
+&i\omega (C_1-C_2)[(V^B_{i,j,X^+}-V^A_{i,j+1,X^+})+(V^B_{i,j,X^+}-V^B_{i,j+1,X^+})],
\end{split}
\end{align}
\begin{align}
\begin{split}
&I^B_{i,j,X^-}\\
=&\frac{V^B_{i,j,X^-}-V^B_{i,j,X^+}}{i\omega L_0}+i\omega C_0 V^B_{i,j,X^-}\\
+&i\omega C[(V^B_{i,j,X^-}-V^A_{i-1,j,Y^-})+(V^B_{i,j,X^-}-V^B_{i-1,j,X^-})\\
+&(V^B_{i,j,X^-}-V^A_{i+1,j,Y^+})+(V^B_{i,j,X^-}-V^B_{i+1,j,X^-})]\\
+&i\omega (C_1+C_2)[(V^B_{i,j,X^-}-V^A_{i,j-1,X^+})+(V^B_{i,j,X^-}-V^B_{i,j-1,X^-})]\\
+&i\omega (C_1-C_2)[(V^B_{i,j,X^-}-V^A_{i,j+1,X^-})+(V^B_{i,j,X^-}-V^B_{i,j+1,X^-})],
\end{split}
\end{align}
\begin{align}
\begin{split}
&I^B_{i,j,Y^+}\\
=&\frac{V^B_{i,j,Y^+}-V^B_{i,j,Y^-}}{i\omega L_0}+i\omega C_0 V^B_{i,j,Y^+}\\
+&i\omega C[(V^B_{i,j,Y^+}-V^A_{i-1,j,X^-})+(V^B_{i,j,Y^+}-V^B_{i-1,j,Y^+})\\
+&(V^B_{i,j,Y^+}-V^A_{i+1,j,X^+})+(V^B_{i,j,Y^+}-V^B_{i+1,j,Y^+})]\\
+&i\omega (C_1+C_2)[(V^B_{i,j,Y^+}-V^A_{i,j-1,Y^-})+(V^B_{i,j,Y^+}-V^B_{i,j-1,Y^+})]\\
+&i\omega (C_1-C_2)[(V^B_{i,j,Y^+}-V^A_{i,j+1,Y^+})+(V^B_{i,j,Y^+}-V^B_{i,j+1,Y^+})],
\end{split}
\end{align}
and
\begin{align}
\begin{split}
&I^B_{i,j,Y^-}\\
=&\frac{V^B_{i,j,Y^-}-V^B_{i,j,Y^+}}{i\omega L_0}+i\omega C_0 V^B_{i,j,Y^-}\\
+&i\omega C[(V^B_{i,j,Y^-}-V^A_{i-1,j,X^+})+(V^B_{i,j,Y^-}-V^B_{i-1,j,Y^-})\\
+&(V^B_{i,j,Y^-}-V^A_{i+1,j,X^-})+(V^B_{i,j,Y^-}-V^B_{i+1,j,Y^-})]\\
+&i\omega (C_1+C_2)[(V^B_{i,j,Y^-}-V^A_{i,j-1,Y^+})+(V^B_{i,j,Y^-}-V^B_{i,j-1,Y^-})]\\
+&i\omega (C_1-C_2)[(V^B_{i,j,Y^-}-V^A_{i,j+1,Y^-})+(V^B_{i,j,Y^-}-V^B_{i,j+1,Y^-})].
\end{split}
\end{align}
We have
\begin{align}
\begin{split}
U^A_{i,j,X}=&\frac{\omega^2 L_0(C_0+8C)}{2}U^A_{i,j,X}\\
&+\frac{\omega^2 L_0 C}{2}(U^A_{i-1,j,X}-U^B_{i-1,j,Y}+U^A_{i+1,j,X}+U^B_{i+1,j,Y})\\
&+\frac{\omega^2 L_0 (C_1+C_2)}{2}(U^A_{i,j-1,X}-U^B_{i,j-1,X})\\
&+\frac{\omega^2 L_0 (C_1-C_2)}{2}(U^A_{i,j+1,X}+U^B_{i,j+1,X}),
\end{split}
\end{align}
\begin{align}
\begin{split}
U^A_{i,j,Y}=&\frac{\omega^2 L_0(C_0+8C)}{2}U^A_{i,j,Y}\\
&+\frac{\omega^2 L_0 C}{2}(U^A_{i-1,j,Y}+U^B_{i-1,j,X}+U^A_{i+1,j,Y}-U^B_{i+1,j,X})\\
&+\frac{\omega^2 L_0 (C_1+C_2)}{2}(U^A_{i,j-1,Y}-U^B_{i,j-1,Y})\\
&+\frac{\omega^2 L_0 (C_1-C_2)}{2}(U^A_{i,j+1,Y}+U^B_{i,j+1,Y}),
\end{split}
\end{align}
\begin{align}
\begin{split}
U^B_{i,j,X}=&\frac{\omega^2 L_0(C_0+8C)}{2}U^B_{i,j,X}\\
&+\frac{\omega^2 L_0 C}{2}(-U^A_{i-1,j,Y}-U^B_{i-1,j,X}+U^A_{i+1,j,Y}-U^B_{i+1,j,X})\\
&+\frac{\omega^2 L_0 (C_1+C_2)}{2}(U^A_{i,j-1,X}-U^B_{i,j-1,X})\\
&+\frac{\omega^2 L_0 (C_1-C_2)}{2}(-U^A_{i,j+1,X}-U^B_{i,j+1,X}),
\end{split}
\end{align}
and
\begin{align}
\begin{split}
U^B_{i,j,Y}=&\frac{\omega^2 L_0(C_0+8C)}{2}U^B_{i,j,Y}\\
&+\frac{\omega^2 L_0 C}{2}(U^A_{i-1,j,X}-U^B_{i-1,j,Y}-U^A_{i+1,j,X}-U^B_{i+1,j,Y})\\
&+\frac{\omega^2 L_0 (C_1+C_2)}{2}(U^A_{i,j-1,Y}-U^B_{i,j-1,Y})\\
&+\frac{\omega^2 L_0 (C_1-C_2)}{2}(-U^A_{i,j+1,Y}-U^B_{i,j+1,Y}).
\end{split}
\end{align}
The voltages of the site $(i,j)$ for A and B read:
\begin{align}
\begin{split}
U^A_{i,j}&=U^A_{i,j,X}+iU^A_{i,j,Y}\\
&=\frac{\omega^2 L_0(C_0+8C)}{2}U^A_{i,j}\\
&+\frac{\omega^2 L_0 C}{2}(U^A_{i-1,j}-iU^B_{i-1,j}+U^A_{i+1,j}+iU^B_{i+1,j})\\
&+\frac{\omega^2 L_0 (C_1+C_2)}{2}(U^A_{i,j-1}-U^B_{i,j-1})\\
&+\frac{\omega^2 L_0 (C_1-C_2)}{2}(U^A_{i,j+1}+U^B_{i,j+1})
\end{split}
\end{align}
\begin{align}
\begin{split}
U^B_{i,j}&=U^B_{i,j,X}+iU^B_{i,j,Y}\\
&=\frac{\omega^2 L_0(C_0+8C)}{2}U^B_{i,j}\\
&+\frac{\omega^2 L_0 C}{2}(-iU^A_{i-1,j}-U^B_{i-1,j}+iU^A_{i+1,j}-U^B_{i+1,j})\\
&+\frac{\omega^2 L_0 (C_1+C_2)}{2}(U^A_{i,j-1}-U^B_{i,j-1})\\
&+\frac{\omega^2 L_0 (C_1-C_2)}{2}(-U^A_{i,j+1}-U^B_{i,j+1}).
\end{split}
\end{align}
Therefore,
\begin{align}
\begin{split}
E=\frac{C_0}{C}+8-\frac{2}{\omega^2 L_0 C}
\end{split}
\end{align}
and
\begin{align}
\begin{split}
f=\frac{1}{2\pi}\frac{1}{\sqrt{L_0C}}\sqrt{\frac{2}{C_0/C+8-E}}.
\end{split}
\end{align}
In the main text, we give the frequency spectrum for the non-Hermitian Chern insulator with $M=-1.5$ and $p/q=1/2$ in Fig. \ref{f6}.

\end{document}